\documentclass[aps,prd,twocolumn,floatfix,preprintnumbers,nofootinbib,superscriptaddress]{revtex4}

\usepackage{amsmath}
\usepackage{graphicx}
\usepackage{textcomp}
\usepackage{amsfonts}
\usepackage{amssymb}
\usepackage{bm}
\usepackage[dvips]{color}
\usepackage{multirow}
\usepackage{booktabs}
\usepackage{epsfig}
\usepackage{subfigure}
\usepackage[utf8]{inputenc}

\begin{document}

\title{The quark orbital angular momentum of ground state octet baryons}

\author{Jing-Feng Li}
\affiliation{School of Physical Science and Technology, Southwest University, Chongqing 400715, China}

\author{Cheng Chen}
\affiliation{Institute of Modern Physics, Chinese Academy of Sciences, Lanzhou 730000, China}
\affiliation{School of Nuclear Science and Technology, University of Chinese Academy of Sciences, Beijing 101408, China} 

\author{Gang Li}~\email{gli@qfnu.edu.cn}
\affiliation{College of Physics and Engineering, Qufu Normal University, Qufu 273165, China}

\author{Chun-Sheng An}~\email{ancs@swu.edu.cn}
\affiliation{School of Physical Science and Technology, Southwest University, Chongqing 400715, China}

\author{Cheng-Rong Deng}~\email{crdeng@swu.edu.cn}
\affiliation{School of Physical Science and Technology, Southwest
University, Chongqing 400715, China}

\author{Ju-Jun Xie}~\email{xiejujun@impcas.ac.cn}
\affiliation{Institute of Modern Physics, Chinese Academy of Sciences, Lanzhou 730000, China} 
\affiliation{School of Nuclear Science and Technology, University of Chinese Academy of Sciences, Beijing 101408, China} 
\affiliation{Lanzhou Center for Theoretical Physics, Key
Laboratory of Theoretical Physics of Gansu Province, Lanzhou University, Lanzhou 730000, China}

\thispagestyle{empty}

\date{\today}

%%%%%%%%%%%%%%%%%%%%%%%%%%%%% Abstract %%%%%%%%%%%%%%%%%%%%%%%%%%%%%

\begin{abstract}

Here we study the quark orbital angular momentum of the ground octet baryons employing an extended chiral constituent quark model, within which the baryon wave functions are taken to be superposition of the traditional $qqq$ and the $qqqq\bar{q}$ higher Fock components. Coupling between the two configurations is estimated using the $^3P_{0}$ quark-antiquark creation mechanism, and the corresponding coupling strength is determined by fitting the sea flavor asymmetry of the nucleon. The obtained numerical results show that the quark angular momentum of the nucleon, $\Sigma$, $\Lambda$ and $\Xi$ hyperons are in the range $0.10$-$0.30$. In addition, the quark angular momentum of all the hyperons are a little bit smaller than that of the nucleon.
And the octet baryons spin fractions taken by the intrinsic quark orbital angular momentum could be up to $60\%$ in present model.

\end{abstract}

\maketitle

%%%%%%%%%%%%%%%%%%%%%%%%%%%%%% Sec. %%%%%%%%%%%%%%%%%%%%%%%%%%%%%%%%%%%

\section{Introduction}
\label{intro}

In the traditional constituent quark models, the ground octet baryons are composed of three quarks that are in their $S$-wave, thus the total spin of the baryons are contributed from only the spin of their quark content, namely, the $SU(2)$ mixed symmetric spin configuration of the three-quark system with total spin $S=1/2$ should account for the total spin of corresponding baryons completely~\cite{Capstick:2000qj,Glozman:1995fu}. However, in the late 1980s, the famous European muon collaboration (EMC) found that the quark spin could take only a small fraction of the proton spin~\cite{Ashman:1987hv,Ashman:1989ig}, which raised up the proton spin crisis. 

Since then intensively experimental measurements on the spin structure of proton have been taken~\cite{Anthony:2000fn,COMPASS:2006mhr,COMPASS:2010wkz,COMPASS:2015mhb,COMPASS:2016jwv,COMPASS:2017hef,Deur:2021klh}, and, on the theoretical side, the spin decomposition of proton in gauge theory was proposed in Refs.~\cite{Ji:1996ek,Balitsky:1997rs,Ji:1997pf,Chen:2008ag}. Accordingly, contributions of the quarks, the gluons and orbital angular momentum (OAM) to the total proton spin have been investigated by lattice QCD~\cite{Aoki:1996pi,Hagler:2007xi,QCDSF:2011aa,Yang:2016plb,Alexandrou:2017oeh,Yamanaka:2018uud,Yang:2019dha,RQCD:2019jai}, chiral perturbation theory~\cite{Chen:2001pva,Dorati:2007bk,Lensky:2014dda,Li:2015exr}, as well as other theoretical approaches~\cite{Brodsky:1988ip,Qing:1997th,Qing:1998at,Brodsky:2000ii,An:2005cj,Brodsky:2006ha,Barquilla-Cano:2006hku,Adamuscin:2007fk,Myhrer:2007cf,Thomas:2008ga,Bijker:2009up,Lorce:2011kd}, for recent reviews on status of the experimental and theoretical investigations on nucleon spin structure, see Refs.~\cite{Kuhn:2008sy,Burkardt:2008jw,Leader:2013jra,Wakamatsu:2014zza,Liu:2015xha,Deur:2018roz,Ji:2020ena}. 

Based on these previous theoretical studies, one may conclude that the intrinsic sea content in baryons should play important roles in the static properties of baryons. On the other hand, the experimentally observed sea flavor asymmetry of proton~\cite{NuSea:2001idv,SeaQuest:2021zxb} also reveals the non-perturbative effects of the intrinsic sea content in baryons. 

Constructively, the proton can be depicted by a meson-cloud picture that the proton is explained to be a neutron core surrounding by pion meson cloud. Furthermore, one can also consider effects of the $K \Lambda$, $K \Sigma$, and $\pi \Delta$ components in proton~\cite{Ericson:1983um,Thomas:1983fh,Henley:1990kw,Brodsky:1996hc}. Within the meson cloud model, spin of proton could be directly decomposed into the spin and orbital angular momentum of quarks, since the meson clouds should be in their $P$-wave states relative to the baryon cores. And the sea flavor asymmetry may be described by the $\pi N$ and $\pi \Delta$ components with appropriate probabilities in proton. Within this picture, one can also explain the intrinsic strange-antistrange quark asymmetry in proton if the $K \Lambda$ and $K \Sigma$ components are taken into account~\cite{Brodsky:1996hc}. Straightforwardly, the meson cloud picture for nucleon can be extended to the other octet baryons~\cite{Shao:2010wq}. 

Alternatively, the extended chiral constituent quark model (E$\chi$CQM), in which the higher Fock components in the baryon's wave function are assumed to be compact pentaquark configurations, was proposed to study the strangeness magnetic moment~\cite{Zou:2005xy}, strangeness form factor~\cite{Riska:2005bh} and strangeness spin of proton~\cite{An:2005cj}, and further developed to investigate the intrinsic sea content and meson-baryon sigma terms of the octet baryons~\cite{An:2012kj,An:2014aea,Duan:2016rkr}. In addition, the experimental data for decays of baryon resonances such as $\Delta(1232)$, $P_{11}(1440)$, $S_{11}(1535)$, $S_{11}(1650)$, $D_{13}(1520)$ and $D_{13}(1700)$~\cite{Li:2005jn,Li:2005jb,Li:2006nm,An:2008xk,An:2009uv,An:2011sb} can be also well reproduced by E$\chi$CQM. Very recently, the E$\chi$CQM has been applied to study the quark OAM of the proton and flavor-dependent axial charges of the ground state octet baryons~\cite{An:2019tld,Wang:2021ild,Qi:2022sus}. It's shown that the singlet axial charge $g_A^{(0)}$, the isovector axial charge $g_A^{(3)}$ and the $SU(3)$ octet axial charge $g_A^{(8)}$ of the octet baryons obtained in the E$\chi$CQM, are consistent with predictions by lattice QCD and chiral perturbation theory, if the model parameters are fixed by fitting the data for $\bar{d}-\bar{u}$ asymmetry of proton~\cite{SeaQuest:2021zxb}. As an direct extension of Ref.~\cite{An:2019tld}, here we investigate the quark OAM of the ground octet baryons, and analyze the spin decomposition of the corresponding baryons in the E$\chi$CQM.

The present manuscript is organized as follows. In Sec.~\ref{frame},
we present the framework which includes the E$\chi$CQM and the formalism for calculations on the OAM in corresponding model, the explicit numerical results and discussions are given in Sec.~\ref{num}. Finally, we give a brief summary of present work in Sec.~\ref{conc}.

%%%%%%%%%%%%%%%%%%%%%%%%%%%%%% Sec. II %%%%%%%%%%%%%%%%%%%%%%%%%%%%%%%%

\section{Framework}
\label{frame}

To investigate the quark orbital angular momentum of the ground octet baryons, here we employ the E$\chi$CQM, in which the wave functions of baryons can be expressed in a general form as:
\begin{equation}
|B\rangle=\frac{1}{\sqrt{\mathcal{N}}}\left(|qqq\rangle+\sum_{i}C_i^{q}|qqq(q\bar{q}),i\rangle\right) 
\label{wfc},
\end{equation}
where the first term just represents the wave functions for the traditional three-quark components of the octet baryons, while the second term denotes the wave functions for the compact five-quark components, with the sum over $i$ runs over all the possible five-quark configurations with a $q\bar{q}$ ($q=u,d,s$) pair that may form considerable higher Fock components in the octet baryons. $C_{i}^q/\sqrt{\mathcal{N}}$ are the corresponding probability amplitudes for the five-quark components with $\mathcal{N}$ being a normalization constant. The seventeen possible $qqq(q\bar{q})$ configurations with $i=1\cdots17$ are shown in Table~\ref{con}, where the flavor, spin, color and orbital wave functions for the four-quark subsystem are denoted by the Young tableaux of the $S_4$ permutation group.

%%%%%%%%%%%%%%%%%%%%%%%% Table I %%%%%%%%%%%%%%%%%%%%%%%%%%%%%%%%%%%%%%%%%%%

{\squeezetable
\begin{table*}[htbp]
\caption{\footnotesize The orbital-flavor-spin configurations for
the four-quark subsystem of the five-quark configurations those may exist as higher Fock components in ground octet baryons.} 
\label{con}
\renewcommand
\tabcolsep{0.10cm}
\renewcommand{\arraystretch}{2}
\scriptsize \vspace{0.7cm}
\begin{tabular}{cccccc}
    \toprule[1.2pt]
    $i$  &  $1$  &  $2$  &  $3$  &  $4$  &  $5$  \\
    Config.&$[31]_{\chi}[4]_{FS}[22]_F[22]_S$&$[31]_{\chi}[31]_{FS}[211]_F[22]_S$&$[31]_{\chi}[31]_{FS}[31]_{F_1}[22]_S$&$[31]_{\chi}[31]_{FS}[31]_{F_2}[22]_S$&$[4]_{\chi}[31]_{FS}[211]_F[22]_S$ \\
\hline

    $i$  &  $6$  &  $7$  &  $8$  &  $9$  &  $10$  \\
    Config.&$[4]_{\chi}[31]_{FS}[31]_{F_1}[22]_S$&$[4]_{\chi}[31]_{FS}[31]_{F_2}[22]_S$&$[31]_{\chi}[4]_{FS}[31]_{F_1}[31]_S$&$[31]_{\chi}[4]_{FS}[31]_{F_2}[31]_S$&$[31]_{\chi}[31]_{FS}[211]_F[31]_S$ \\
\hline

    $i$  &  $11$  &  $12$  &  $13$  &  $14$  &  $15$  \\
    Config.&$[31]_{\chi}[31]_{FS}[22]_F[31]_S$&$[31]_{\chi}[31]_{FS}[31]_{F_1}[31]_S$&$[31]_{\chi}[31]_{FS}[31]_{F_2}[31]_S$&$[4]_{\chi}[31]_{FS}[211]_F[31]_S$&$[4]_{\chi}[31]_{FS}[22]_F[31]_S$ \\
\hline

    $i$  &  $16$  &  $17$  &    &   &    \\
    Config.&$[4]_{\chi}[31]_{FS}[31]_{F_1}[31]_S$&$[4]_{\chi}[31]_{FS}[31]_{F_2}[31]_S$&&& \\
 \bottomrule[1.2pt]
\end{tabular}

\end{table*}
}
%%%%%%%%%%%%%%%%%%%%%%%%%%%%%%%%%%%%%%%%%%%%%%%%%%%%%%%%%%%%%%%%%%%%%%%%%%%%%%%%

As we can see in Table~\ref{con}, spin wave functions of the four-quark subsystem in configurations with $i=1\cdots7$ are $[22]_S$, which leads to the total spin $S_4=0$ for the four-quark subsystem, thus one only need to consider the combination of the quark (antiquark) OAM and spin of the antiquark to get the total spin of a given baryon. And a general form for the wave functions of these configurations can be expressed as
\begin{widetext}
\begin{eqnarray}
|B,i=1\cdots7\rangle_{5q} &=& \sum_{ijkln}\sum_{ab}\sum_{m\bar{s}_z}C^{\frac{1}{2},\uparrow}_{1,m;\frac{1}{2},\bar{s}_z}C^{[1^4]}_{[31]_{\chi FS}^k;[211]_C^{\bar{k}}}C^{[31]_{\chi FS}^{k}}_{[O]_\chi^i;[FS]_{FS}^j}
C^{[FS]_{FS}^{j}}_{[F]_F^l;[22]_{S}^n}C_{a,b}^{[2^{3}]_{C}}|[211]_C^{\bar{k}}(a)\rangle|[11]_{C,\bar{q}}(b)\rangle|I,I_{3}\rangle^{[F]_{F}^{l}}
\nonumber\\
&&|1,m\rangle^{[O]_\chi^i}|[22]_S^n\rangle|\bar{\chi},\bar{s}_z\rangle\phi(\{\vec{r}_q\}),
\label{22s}
\end{eqnarray}
\end{widetext}
with the coefficients $C^{[\cdots]}_{[\cdots][\cdots]}$ represent the CG coefficients of
the $S_{4}$ permutation group, and $C^{\frac{1}{2},\uparrow}_{1,m;\frac{1}{2},\bar{s}_z}$ the CG coefficients for the combination of the quark (antiquark) OAM and spin of the antiquark to form a spin $|1/2,+1/2\rangle$ baryon state. And the explicit flavor, orbital, spin and color wave functions for the presently considered five-quark configurations have been given in Refs.~\cite{Qi:2022sus}.

While for the five-quark configurations with $i=8\cdots17$ shown in Table~\ref{con}, the spin wave function of the four-quark subsystem is $[31]_S$ which results in the total spin $S_4=1$ for the four-quark subsystem. Accordingly, we have to take into account the combination of the spin of both the four-quark and the antiquark and the quark (antiquark) OAM to get the total spin of the octet baryon. One should note that combination of the spin for four-quark subsystem $S_4=1$ and the quark OAM $L=1$ will lead to $J=L\oplus S_4=0$,~$1$~or~$2$, and the former two $J$ those could form the total spin $S_B=1/2$ of the presently studied octet baryons, when combine to the spin of the antiquark $S_{\bar{q}}=1/2$ are applicable. Hereafter, we denote these two cases of wave functions as Set I and Set II, respectively. 

The general forms for wave functions of the five-quark configurations with $i=8\cdots17$ in these two cases are then given by 
\begin{widetext}
\begin{eqnarray}
|B,i=8\cdots17\rangle_{5q}^{\mathrm{I}} &=&  \sum_{ijkln}\sum_{ab}\sum_{ms_z}C^{00}_{1,m;1,s_z}C^{[1^4]}_{[31]_{\chi FS}^k;[211]_C^{\bar{k}}}C^{[31]_{\chi FS}^{k}}_{[O]^\chi_i;[FS]_{FS}^j}
 C^{[FS]_{FS}^{j}}_{[F]_F^l;[31]_{S}^n}C_{a,b}^{[2^{3}]_{C}}|[211]_C^{\bar{k}}(a)\rangle|[11]_{C,\bar{q}}(b)\rangle|I,I_3\rangle^{[F]_{F}^{l}} \nonumber \\
&&
|1,m\rangle^{[O]_\chi^i}|[31]_S^n,s_z\rangle|\bar{\chi},\bar{s}_z\rangle\phi(\{\vec{r}_q\})\,,
\label{31s0}
\end{eqnarray}
and
\begin{eqnarray}
|B,i=8\cdots17\rangle_{5q}^{\mathrm{II}} &=& 
\sum_{ijkln}\sum_{ab}\sum_{J_z\bar{s}_z}\sum_{ms_z}
C^{\frac{1}{2},\frac{1}{2}}_{1,J_z;\frac{1}{2},\bar{s}_z}C^{1,J_z}_{1,m;1,s_z}C^{[1^4]}_{[31]_{\chi FS}^k;[211]_C^{\bar{k}}}
 C^{[31]_{\chi FS}^{k}}_{[O]_\chi^i;[FS]_{FS}^j}C^{[FS]_{FS}^{j}}_{[F]_F^l;[31]_{S}^n}C_{a,b}^{[2^{3}]_{C}}|[211]_C^{\bar{k}}(a)\rangle|[11]_{C,\bar{q}}(b)\rangle \nonumber \\
&&  |I,I_3\rangle^{[F]_{F}^{l}}
|1,m\rangle^{[O]^\chi_i}|[31]^S_n,s_z\rangle|\bar{\chi},\bar{s}_z\rangle\phi(\{\vec{r}_q\})\,.
\label{31s1}
\end{eqnarray}
\end{widetext}

Next we turn to the coefficients $C_i^q$ in Eq.~\eqref{wfc}. Generally, to get the probability amplitude for a five-quark component in a given baryon, one has to evaluate energy of the five-quark configuration $E_i^q$ and its coupling to the three-quark component in the corresponding baryon. 

In present work, the energy $E_i^q$ for a given five-quark configuration is estimated using the chiral constituent quark model, within which the hyperfine interaction between quarks is~\cite{Glozman:1995fu}
\begin{eqnarray}
H_{hyp} &=& -\sum_{i<j}\delta(r_{ij})\vec{\sigma}_i\cdot
\vec{\sigma}_j\Bigg[\sum_{a=1}^3V_\pi (r_{ij})\lambda^a_i\lambda^a_j
\nonumber\\
&& +\sum_{a=4}^7 V_K (r_{ij})\lambda^a_i\lambda^a_j +V_\eta
(r_{ij})\lambda^8_i\lambda^8_j\Bigg]\, ,
\end{eqnarray}
where $\vec{\sigma}_{i(j)}$ and $\lambda^a_{i(j)}$ are the Pauli and flavor $SU(3)$ Gell-Mann matrices acting on the $i(j)$-th quark, and $V_M(r_{ij})$ denotes the potential for exchanging a $M$-meson. Calculations on the matrix elements $\langle qqq(q\bar{q}),i|H_{hyp}|qqq(q\bar{q}),i\rangle$ will lead to the following common factors
\begin{equation}
    P^{M}_l=\langle lm|\delta(r_{ij})V_M(r_{ij})|lm\rangle\,,
\end{equation}
with $|lm\rangle$ the spatial wave function with OAM quantum number $l$. Here we just take the empirical values for $P^{M}_l$ those could very well reproduce the spectroscopy of light and strange baryons~\cite{Glozman:1995fu}:
\begin{eqnarray}
& P^{\pi}_{0}=29~\mathrm{MeV},~P^{K}_{0}=20~\mathrm{MeV},~P^{s\bar{s}}_{0}=14~\mathrm{MeV}\,,\nonumber\\
& P^{\pi}_{1}=45~\mathrm{MeV},~P^{K}_{1}=30~\mathrm{MeV},~P^{s\bar{s}}_{1}=20~\mathrm{MeV}\,.
\label{plm}
\end{eqnarray}
Then the energy $E_i^q$ for the five-quark configurations listed in Table~\ref{con} can be calculated by
\begin{equation}
    E_{i}^q=E_{0}+\langle qqq(q\bar{q}),i|H_{hyp}|qqq(q\bar{q}),i\rangle + n_i^s\delta m\,,
\end{equation}
where $E_{0}$ is a degenerated energy for all the studied configurations when the hyperfine interaction between quarks and the flavor $SU(3)$ breaking effects are not taken into account, and $\delta_m$ and $n_i^s$ denote the mass difference of the light and strange quarks and number of strange quarks in the corresponding five-quark system, respectively. Here both $E_{0}$ and $\delta_m$ are taken to be the empirical values~\cite{An:2012kj}:
\begin{equation}
    E_{0}=2127~\mathrm{MeV},~~~\delta_m=120~\mathrm{MeV}\,.
    \label{e0dm}
\end{equation}
%

%%%%%%%%%%%%%%%%%%%%%%%%%%%%%%%%%%  FIG. I %%%%%%%%%%%%%%%%%%%%%%%%%%%%%%%%%%
\begin{figure}[b]
\begin{center}
\includegraphics[scale=0.55]{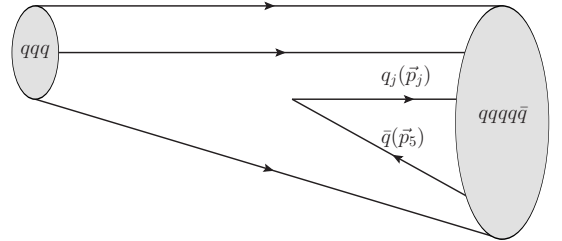}
\end{center}
{\caption{Transition $qqq\rightarrow qqqq\bar{q}$ caused by a
quark-antiquark pair creation in a baryon via the $^{3}P_{0}$
mechanism.} 
\label{3to5}}
\end{figure}
%%%%%%%%%%%%%%%%%%%%%%%%%%%%%%%%%%%%%%%%%%%%%%%%%%%%%%%%%%%%%%%%%%%%%%%%%%%%%%

To calculate the transition coupling matrix elements between the $qqq$ and $qqqq\bar{q}$ components, here we adopt a widely accepted $^3P_0$ quark-antiquark creation mechanism~\cite{LeYaouanc:1972vsx,LeYaouanc:1973ldf}, which has been used to study the intrinsic sea content of nucleon~\cite{An:2012kj}. Explicitly, as depicted in Fig.~\ref{3to5}, the three initial quarks go as spectators, and a quark-antiquark pair with quantum number $J^{P}=0^+$ is created in the vacuum, and then form the final five-quark system. The operator for this kind of transition can be written as
\begin{eqnarray}
\hat{T} &=& -\gamma \sum_{j=1,4}
\mathcal{F}_{j,5}^{00}\mathcal{C}_{j,5}^{00}\mathcal{C}_{OFSC}\sum_{m}\langle
1,m;1,-m|00\rangle \times \nonumber \\
&&
\chi_{j,5}^{1,m}\mathcal{Y}_{j,5}^{1,-m}(\vec{p}_j-\vec{p}_5)b^{\dagger}(\vec{p}_j)d^{\dagger}(\vec{p}_5)
\, , \label{3p0}
\end{eqnarray}
where $\gamma$ is an dimensionless transition coupling constant, $\mathcal{F}_{j,5}^{00}\text{ and
}\mathcal{C}_{j,5}^{00}$ are the flavor and color singlet of the
created quark-antiquark pair $q_j\bar{q}_5$, $\chi_{j,5}^{1,m}\text{
and }\mathcal{Y}_{j,5}^{1,-m}$ are the total spin $S_{q\bar{q}}=1$ and relative
orbital $P-$ state of the created quark-antiquark system, the
operator $\mathcal{C}_{OFSC}$ is to calculate the overlap factor
between the residual three-quark configuration in the five-quark
component and the valence three-quark component, finally,
$b^{\dagger}(\vec{p}_j),d^{\dagger}(\vec{p}_5)$ are the quark and
antiquark creation operators.

Then, one can calculate the probability amplitude for a five-quark configuration $ |qqq(q\bar{q}),i\rangle$ in a given baryon $B$ by the following equation:
\begin{equation}
C_i^q=\frac{\langle qqq(q\bar{q}),i|\hat{T}|qqq\rangle}{M_B-E_i^q}\,,
 \label{coe}
\end{equation}
here $M_B$ denote the physical mass of the baryon $B$~\cite{ParticleDataGroup:2020ssz}.

Explicit calculations on the transition matrix elements $\langle qqq(q\bar{q}),i|\hat{T}|qqq\rangle$ between all the five-quark configurations shown in Table~\ref{con} and the $qqq$ components in the presently studied octet baryons will result in a common factor $\mathcal{V}$, namely,
\begin{equation}
    \langle qqq(q\bar{q}),i|\hat{T}|qqq\rangle=\mathcal{T}_i\mathcal{V}\,,
\end{equation}
the coefficient $\mathcal{T}_i$ can be obtained directly by the given wave function of the $i$-th five-quark configuration, and $\mathcal{V}$ depends on the $^3P_0$ transition coupling constant $\gamma$ and parameters of explicit spatial wave functions determined by a given quark confinement potential.

To reduce free model parameters, here we just fix $\mathcal{V}$ by fitting experimental data for the intrinsic sea flavor asymmetry of the proton~\cite{SeaQuest:2021zxb}
\begin{equation}
    I_a=\bar{d}-\bar{u} = \int_{0}^{1} [\bar{d}_p(x)-\bar{u}_p(x)] dx = 0.118 \pm 0.012\,.
    \label{datasea}
\end{equation}
In present model, the $\bar{d}-\bar{u}$ asymmetry can be obtained by
\begin{eqnarray}
I_a &=&\bar{d}-\bar{u}\nonumber\\
&=& \frac{1}{\mathcal{N}}
\Bigg\{ \Big ( \frac{T_{1}^{l \bar{l}}}{M_p-E_1^l}\Big )^2 +
\Big ( \frac{T_{11}^{l \bar{l}}}{M_p-E_{11}^l}\Big )^2 +
\Big ( \frac{T_{15}^{l \bar{l}}}{M_p-E_{15}^l}\Big )^2\nonumber\\
&-& \frac{1}{3} \bigg[ \Big (\frac{T_3^{l \bar{l}}}{M_p-E_3^l}\Big )^2 +
\Big (\frac{T_6^{l \bar{l}}}{M_p-E_6^l})\Big )^2 +
\Big (\frac{T_8^{l \bar{l}}}{M_p-E_8^l}\Big )^2 \nonumber\\
&+& \Big ( \frac{T_{12}^{l \bar{l}}}{M_p-E_{12}^l}\Big )^2 +
\Big ( \frac{T_{16}^{l \bar{l}}}{M_p-E_{16}^l} \Big )^2\bigg] \Bigg\}\,,
\label{Asym}
\end{eqnarray}
where $T_{i}^{l \bar{l}}$ denotes the transition coupling matrix element between the $i$-th five-quark configuration in Table~\ref{con} with a light quark-antiquark pair $l\bar{l}$ and the three-quark component in proton:
\begin{equation}
    T_{i}^{l \bar{l}}=\langle uud(l\bar{l}),i|\hat{T}|uud\rangle\,,
\end{equation}
and $E_{i}^l$ is energy of the corresponding five-quark configuration with a light quark-antiquark pair. 

Accordingly, the involved parameters in present model are the coupling strengths $P_{l}^{M}$ (six involved ones in total), the degenerated energy $E_0$ for all the five-quark configurations given in Table~\ref{con}, mass difference of the light and strange quark $\delta_m$, and the common factor $\mathcal{V}$ for the transition matrix elements $\langle\hat{T}\rangle$. As shown in Eqs.~(\ref{plm}) and~(\ref{e0dm}), the former eight parameters are taken to be the empirical values used in the literature. And by fitting the sea flavor asymmetry of proton in Eq.~(\ref{datasea}), one can get the following values of $\mathcal{V}$:
\begin{eqnarray}
\mathcal{V}_{\rm I}=570\pm 46~~{\rm MeV}\,,\\
\mathcal{V}_{\rm II}=697\pm {80}~~{\rm MeV}\,,
\end{eqnarray}
for the two different sets of wave functions using in present model, respectively.

Note that one could also consider the higher Fock components $qqq(q\bar{q})^2$ in the baryons. Energies of these kinds of components those can couple to the ground state octet baryons may be about $600-900$~MeV higher than the five-quark components, taking into account the two quark-antiquark pairs creations, a rough estimation shows that the probability amplitudes of the $qqq(q\bar{q})^2$ components are about $1/5$ of those of the presently considered five-quark components. 

Finally, within the E$\chi$CQM, the quark OAM of the octet baryons can be calculated by 
\begin{equation}
L_f=\langle B|\hat{L}_{fz}|B\rangle=\frac{(C_i^q)^2\langle qqq(q\bar{q},i)|\hat{L}_{fz}|qqq(q\bar{q},i)\rangle}{\mathcal{N}}\,,
\label{lqo}
\end{equation}
with the operator $\hat{L}_{fz}$ defined by
\begin{equation}
\hat{L}_{fz}=\sum_{f}(\hat{l}_{f}+\hat{l}_{\bar{f}})_{z}\,,
\end{equation}
where $\hat{l}_{f}$ and $\hat{l}_{\bar{f}}$ are the OAM operators for the quark and antiquark with flavor $f$, respectively, and
the sum runs over the flavors $u$, $d$ and $s$.

%%%%%%%%%%%%%%%%%%%%%%%%%%%%%%%%%%%%%%% Sec. III  %%%%%%%%%%%%%%%%%%%%%%%%%%%%%

\section{Numerical results and discussions}
\label{num}

%%%%%%%%%%%%%%%%%%%%%%%%%%%%%% Figure 2 %%%%%%%%%%%%%%%%%%%%%%%%%%%%%%%%%%%%%%
\begin{figure}[t]
\begin{center}
\includegraphics[scale=0.34]{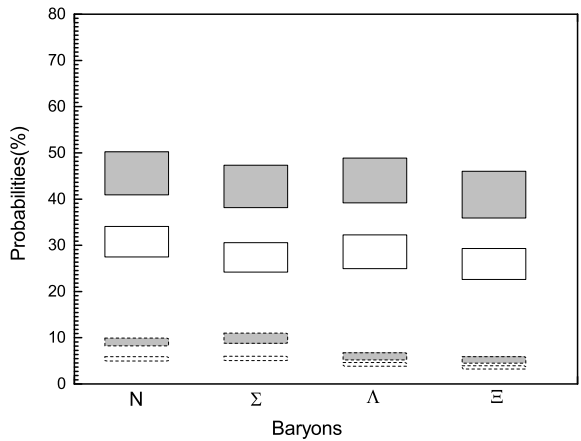}
\end{center}
{\caption{The probabilities of the five-quark components in the octet baryons, the boxes filled by white and grey colors are results obtained in Sets I and II, respectively, and the boxes with solid line border are results for the five-quark components with a light quark-antiquark pair, while those with dash line border are results for the five-quark components with a strange quark-antiquark pair. The rectangle height shows the uncertainty.} 
\label{p5q}}
\end{figure}
%%%%%%%%%%%%%%%%%%%%%%%%%%%%%%%%%%%%%%%%%%%%%%%%%%%%%%%%%%%%%%%%%%%%%%%%%%%%%%%

%%%%%%%%%%%%%%%%%%%%%%%%%%%%%%%%%%%%%%%%  Table II %%%%%%%%%%%%%%%%%%%%%%%%%%%%
\begin{table*}[ht]
\caption{The quark OAM of the octet baryons. the upper and lower panels denoted by Set I and Set II are the numerical results obtained using the two sets of wave functions, respectively.}
\label{lqob}
\renewcommand
\tabcolsep{0.50cm}
\renewcommand{\arraystretch}{2.0}
\begin{tabular}{ccccccc}
\hline\hline
  
   & & $p$ &  $\Sigma^{+}$ & $\Sigma^0$ &  $\Lambda$ & $\Xi^{0}$  \\ 
\hline

\multirow{3}{*}{Set I} & $L_u$ & $0.080(09)$  & $0.063(07)$ &$0.050(06)$ &$0.049(06)$ & $0.041(05)$  \\
                       & $L_d$ & $0.063(07)$  & $0.037(04)$& $0.050(06)$& $0.049(06)$& $0.031(03)$  \\
                       & $L_s$ & $0.014(02)$  & $0.029(03)$& $0.029(03)$&$0.040(05)$ & $0.055(06)$ \\
                       \hline
                       
\multirow{3}{*}{Set II} & $L_u$ &$0.136(15)$ &$0.114(13)$ &$0.091(10)$ &$0.089(10)$ & $0.078(09)$ \\
                       & $L_d$ &$0.105(11)$  &$0.068(09)$ &$0.091(10)$ &$0.089(10)$ &$0.065(08)$ \\
                       & $L_s$ &$0.026(04)$ & $0.060(08)$&$0.060(08)$ & $0.067(08)$&$0.085(11)$ \\

\hline\hline

\end{tabular}
\end{table*}
%%%%%%%%%%%%%%%%%%%%%%%%%%%%%%%%%%%%%%%%%%%%%%%%%%%%%%%%%%%%%%%%%%%%%%%%%%%%%%%%

\subsection{The quark OAM of the octet baryons}

Within the framework of the E$\chi$CQM as shown in Sec.~\ref{frame}, one can calculate the probabilities of all the possible five-quark components in the octet baryons, here we show the probabilities of the five-quark components with light and strange quark-antiquark pairs in each octet baryon in Fig.~\ref{p5q}. As discussed explicitly in~\cite{Qi:2022sus}, the obtained baryon wave functions in the both the Set I and Set II could result in the meson-baryon sigma terms consistent with the predictions by other theoretical approaches, although the probabilities for the five-quark components in the two Sets are very different.
And one may note that the upper limit of probabilities for the intrinsic five-quark components obtained in Set II of present model are larger than $50\%$, while these values are very close to those predicted in Refs.~\cite{Chang:2011vx,Chang:2011du,Chang:2014jba}, where the model proposed by Brodsky, Hoyer, Peterson and Sakai~\cite{Brodsky:1980pb} was employed.

\begin{figure}[hbtp]
\begin{center}
\includegraphics[scale=0.32]{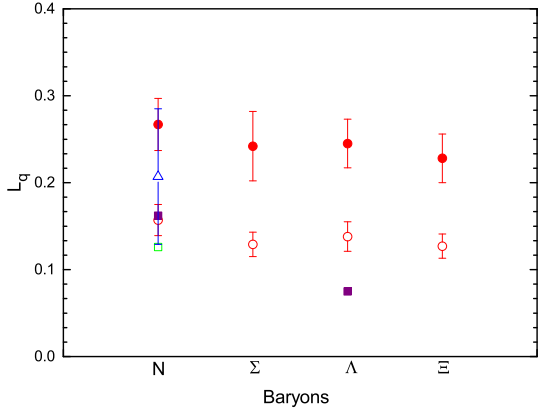}
\end{center}
{\caption{The total quark OAM of all flavors for the octet baryons. The present results in Sets I and II are shown by (red) hollow and solid circles, respectively. And the (blue) triangle, (purple) solid square and (green) hollow square are results predicted by lattice QCD~\cite{Alexandrou:2017oeh}, unquenched quark model~\cite{Bijker:2009up} and light cone constituent quark model~\cite{Lorce:2011kd}, respectively.} 
\label{lq}}
\end{figure}

Using the wave functions shown in Eqs.~(\ref{22s}), (\ref{31s0}) and (\ref{31s1}) with explicit $C_i^q$ calculated by Eq.~(\ref{coe}), one can directly calculate the quark OAM of the octet baryons by Eq.~(\ref{lqo}). The corresponding numerical results for $L_u$, $L_d$ and $L_s$ of the proton, $\Sigma^{+}$, $\Sigma^0$, $\Lambda$ and $\Xi^0$ baryons are presented in Table~\ref{lqob}. And the flavor-dependent quark OAM of neutron, $\Sigma^-$ and $\Xi^-$ can be obtained directly using the $SU(2)$ isospin symmetry, which yields 
\begin{eqnarray}
&L_{u(d)}(n)=L_{d(u)}(p)\,,~~~L_{s}(n)=L_{s}(p)\,,\\
&L_{u(d)}(\Sigma^-)=L_{d(u)}(\Sigma^+)\,,~~~L_{s}(\Sigma^-)=L_{s}(\Sigma^+)\,,\\
&L_{u(d)}(\Xi^-)=L_{d(u)}(\Xi^0)\,,~~~L_{s}(\Xi^-)=L_{s}(\Xi^0)\,.
\end{eqnarray}
And in Fig.~\ref{lq}, we depict the presently obtained total quark OAM $L_q$ of all flavors compared to the results predicted by lattice QCD~\cite{Alexandrou:2017oeh}, unquenched quark model~\cite{Bijker:2009up} and light cone constituent quark model~\cite{Lorce:2011kd}. 

Since all the valence quarks in the octet baryons are in their $S$-wave, the quark OAM should be only contributed from the five-quark higher Fock components of baryons. Therefore, the quark OAM should be sensitive to the probabilities of the five-quark components. Therefore, as we can see in Table~\ref{lqob}, the results for the quark OAM obtained in Set II model are about $1.5-2$ times of those in Set I model. On the other hand, in Set I, the five-quark configurations with $i=8\cdots17$ don't contribute to the quark OAM of proton, as discussed in Ref.~\cite{An:2019tld}, but it's not true for Set II. Consequently, the relationship between the flavor sea asymmetry and total quark OAM of proton in Set II is
\begin{equation}
    L_q\approx 9/4 I_a\,,
\end{equation}
instead of the $\sim4/3I_a$ in Set I obtained in Ref.~\cite{An:2019tld}.

%%%%%%%%%%%%%%%%%%%%%%%%%%%%%%%% Figure 3 %%%%%%%%%%%%%%%%%%%%%%%%%%%%%%%%%%%%%
%%%%%%%%%%%%%%%%%%%%%%%%%%%%%%%%%%%%%%%%%%%%%%%%%%%%%%%%%%%%%%%%%%%%%%%%%%%%%%%

For the nucleon, in both Sets I and II, the up and down quarks OAM are comparable to each other, this is very different from the lattice QCD predictions in Ref.~\cite{Alexandrou:2017oeh}, where the obtained light quark OAM are $L_u=-0.107(40)$ and $L_d=0.247(38)$, while the total quark OAM of all flavors in Ref.~\cite{Alexandrou:2017oeh} is in the same range of the presently obtained results, as shown in Fig.~\ref{lq}. In addition, the presently obtained total quark OAM is in consistent with the predictions by unquenched quark model~\cite{Bijker:2009up} and light cone constituent quark model~\cite{Lorce:2011kd}.

For the other octet baryons, the light quarks OAM of the hyperons are smaller than that of nucleon, while the strange quark OAM of hyperons are $2-4$ times of that of nucleon. Obviously, this is because of the more strange quark contents in hyperons than in nucleon. In Ref.~\cite{Bijker:2009up}, the total quark OAM of $\Lambda$ was studied employing the unquenched quark model by considering the effects of the quark-antiquark pairs created via the $^3P_0$ mechanism, and the predicted $L_q$ of $\Lambda$ hyperon is $0.075$, that is smaller than the presently obtained results in both Sets I and II. 

\subsection{The spin decomposition of the octet baryons}

It's also very interesting to investigate the spin decomposition of the octet baryons. In present model, the spin and OAM of gluons are not involved, accordingly, one can decompose the octet baryons spin by
\begin{equation}
    S_B=1/2=\sum_f \Big[\Big(\Delta f^{val}+\Delta f^{sea}\Big)/2+L_f^{sea}\Big]\,,
\end{equation}
where $\Delta f^{val}/2$ and $\Delta f^{sea}/2$ are the flavor-dependent valence and intrinsic sea quarks spin, respectively, and $L_f^{sea}$ the intrinsic sea quark OAM which should be the same as those numerical results shown in Table~\ref{lqob}, and the sum over $f$ runs over the three flavors $u$, $d$ and $s$. We present our numerical results of Sets I and II obtained by limiting the sea flavor asymmetry in proton to be the experimentally measured central value $I_a=\bar{d}-\bar{u}=0.118$~\cite{SeaQuest:2021zxb} in Table~\ref{spind}, compared to the results in the traditional three-quark constituent quark model. 

%%%%%%%%%%%%%%%%%%%%%%%%%%%%%%%%%%%%%%%%  Table III %%%%%%%%%%%%%%%%%%%%%%%%%%%%
\begin{table*}[ht]
\caption{Spin decomposition of the octet baryons. The three panels from left to right denoted by Traditional three-quark CQM, E$\chi$CQM Set I and E$\chi$CQM Set II are results in the traditional three-quark constituent quark model and the presently obtained results in Sets I and II, respectively. The last three rows denoted by $\mathcal{C}^{val}_\Delta$, $\mathcal{C}^{sea}_\Delta$, $\mathcal{C}^{\Sigma}_\Delta$ and $\mathcal{C}^{sea}_L$ are baryon spin fractions carried by the valence quark spin, intrinsic sea quark spin, total (valence and sea) quark spin and intrinsic sea quark OAM, respectively, in unit of $\%$.}
\label{spind}
\renewcommand
\tabcolsep{0.12cm}
\renewcommand{\arraystretch}{2.0}
\begin{tabular}{cccccccccccccccc}
\hline\hline
  &  \multicolumn{5}{c}{Traditional three-quark CQM}  & \multicolumn{5}{c}{E$\chi$CQM Set I} & \multicolumn{5}{c}{E$\chi$CQM Set II}\\
  \hline
    & $p$ & $\Sigma^{+}$ & $\Sigma^0$ & $\Lambda$ & $\Xi^{0}$ & $p$ & $\Sigma^{+}$ & $\Sigma^0$ & $\Lambda$ & $\Xi^{0}$ & $p$ & $\Sigma^{+}$ & $\Sigma^0$ & $\Lambda$ & $\Xi^{0}$ \\ 
\hline

$\Delta u^{val}$ & $4/3$ &  $4/3$ & $2/3$ &  $0$ & $-1/3$ & $0.839$ & $0.883$ & $0.442$ &$0$ & $-0.233$ & $0.606$ & $0.638$ & $0.319$ & $0$ & $-0.177$
\\

$\Delta d^{val}$ & $-1/3$ &  $0$ & $2/3$ &  $0$ & $0$ & $-0.210$ & $0$ & $0.442$ & $0$ & $0$ & $-0.152$ & $0$ & $0.319$ & $0$ & $0$
\\

$\Delta s^{val}$ & $0$ &  $-1/3$ & $-1/3$ &  $-1/3$ & $4/3$ & $0$ & $-0.221$ & $-0.221$ & $0.671$ & $0.933$ & $0$ & $-0.160$ & $-0.160$ & $0.497$ & $0.708$
\\
\hline

$\Delta u^{sea}$ & $0$  &  $0$ & $0$ & $0$ & $0$ & $0.044$ & $0.039$ & $0.031$ & $0.026$ & $0.018$ & $0.104$ & $0.124$ & $0.029$ & $-0.020$ & $-0.017$
\\

$\Delta d^{sea}$ & $0$  &  $0$ & $0$ & $0$ & $0$ & $-0.003$ & $0.023$ & $0.031$ & $0.026$ & $0.030$ & $-0.073$ & $-0.066$ & $0.029$ & $-0.020$ & $-0.059$
\\

$\Delta s^{sea}$ & $0$  &  $0$ & $0$ & $0$ & $0$ & $0.015$ & $0.016$ & $0.016$ & $0.003$ & $-0.003$ & $-0.020$ & $-0.020$ & $-0.020$ & $0.054$ & $0.090$
\\

\hline

$L_u^{sea}$ & $0$  &  $0$ & $0$ & $0$ & $0$ & $0.080$ & $0.063$ & $0.050$ & $0.049$ & $0.041$ & $0.136$ & $0.114$ & $0.091$ & $0.089$ & $0.078$
\\

$L_d^{sea}$ & $0$  &  $0$ & $0$ & $0$ & $0$ & $0.063$ & $0.037$ & $0.050$ & $0.049$ & $0.031$ & $0.105$  & $0.068$ & $0.091$ & $0.089$ & $0.065$
\\

$L_s^{sea}$ & $0$  &  $0$ & $0$ & $0$ & $0$ & $0.014$ & $0.029$ & $0.029$ & $0.040$ & $0.055$ & $0.026$ & $0.060$ & $0.060$ & $0.067$ & $0.085$
\\

\hline

$\mathcal{C}^{val}_\Delta$ ($\%$) &  $100$ & $100$ & $100$ & $100$ & $100$ &  $63$ & $66$ & $66$ & $67$ & $70$ & $46$ & $48$ & $48$ & $50$ & $53$
\\

$\mathcal{C}^{sea}_\Delta$ ($\%$) &  $0$ & $0$ & $0$ & $0$ & $0$ &  $6$ & $8$ & $8$ & $5$ & $5$ & $1$ & $4$ & $4$ & $1$ & $1$
\\

$\mathcal{C}^{\Sigma}_\Delta$ ($\%$) &  $100$ & $100$ & $100$ & $100$ & $100$ &  $69$ & $74$ & $74$ & $72$ & $75$ & $47$ & $52$ & $52$ & $51$ & $54$
\\

$\mathcal{C}^{sea}_L$ ($\%$) &  $0$ & $0$ & $0$ & $0$ & $0$ &  $31$ & $26$ & $26$ & $28$ & $25$ & $53$ & $48$ & $48$ & $49$ & $46$
\\

\hline\hline

\end{tabular}
\end{table*}
%%%%%%%%%%%%%%%%%%%%%%%%%%%%%%%%%%%%%%%%%%%%%%%%%%%%%%%%%%%%%%%%%%%%%%%%%%%%%%%%

As shown in Table~\ref{spind}, in the traditional three-quark constituent quark model, only spin of the valence up and down quarks contribute to the nucleon spin. While in the presently employed E$\chi$CQM, spin of the valence and intrinsic sea quarks contribute $\sim63\%$ and $6\%$ of the nucleon spin, respectively, and the nucleon spin arisen from the intrinsic sea quark OAM is $31\%$, if the wave functions of Set I are adopted. This is in good agreement with the results in~\cite{Bijker:2009up} obtained by the unquenched quark model.
While if we use the wave functions of Set II, the nucleon spin contributed from the valence quark spin is $46\%$, and the intrinsic sea quark spin and OAM should contribute $1\%$ and $53\%$ to the nucleon spin, respectively. In addition, if the uncertainty in present model caused by fitting the sea flavor asymmetry data are taken into account, contributions from the quark OAM to the nucleon spin could be up to $60\%$.  

Compared to the experimental data, i.e. the measured negative value for spin of strange quark in proton $\Delta s$, and the proton spin fraction carried by quark spin $\Sigma=0.33(10)$~\cite{Anthony:2000fn,COMPASS:2006mhr,COMPASS:2010wkz,COMPASS:2015mhb,COMPASS:2016jwv,COMPASS:2017hef,Deur:2021klh}, one may conclude that the present Set II of wave functions should be more favorable than Set I, although the value $\mathcal{C}^{\Sigma}_\Delta=47\%$ for the quark spin fraction of proton obtained in Set II model is still larger than experimental data. 

The presently obtained quark spin fraction of the proton in the Set II model is very close to the numerical result $0.46$ predicted in~\cite{Barquilla-Cano:2006hku} with contributions of only the one-body axial current. It's shown that the simple additive constituent quark model should violate the partial conservation of the axial current (PCAC) condition~\cite{Barquilla-Cano:2002btm,Barquilla-Cano:2003qmv}, so one has to take into account contributions of two-body axial exchange currents, which could deduce the one-body axial current result by $\sim40\%$~\cite{Barquilla-Cano:2006hku}, and the obtained quark spin fraction of proton including contributions from total currents is consistent with the experimental data.    

In Ref.~\cite{Myhrer:2007cf}, the authors employed a relativistic constituent quark model, and took into account the one-gluon-exchange effects and the pion cloud contributions, which resulted in that the quark OAM could contribute about $62\%$ to the nucleon spin. This value is close to the presently obtained upper limit for the quark OAM in the Set II model.

For the other octet baryons, as shown in Table~\ref{spind}, the baryon spin arisen from the intrinsic sea quark spin should be less than $10\%$, while contributions of the intrinsic sea quark OAM to the baryon spin are about $25-28\%$ and $46-49\%$ using the wave functions of Set I and Set II, respectively. And one should note that the contributions of the valence quark to spin of $\Sigma$, $\Lambda$ and $\Xi$ hyperons in the present Set I model are $66\%$, $67\%$ and $70\%$, respectively, while in Set II model, the corresponding contributions are $48\%$, $50\%$ and $53\%$, respectively. 

In~\cite{Buchmann:2010pd,Buchmann:2014ima}, the quark spin contribution to the total angular momentum of flavor octet and decuplet ground state baryons are studied using a spin-flavor symmetry based parametrization method of quantum
chromodynamics. The results for the quark spin fraction of the ground state octet baryons in Ref.~\cite{Buchmann:2010pd} is $0.35(12)$, and the value obtained in Ref.~\cite{Buchmann:2014ima} is $0.41(12)$, which values are in general consistent with the presently obtained results.

In Ref.~\cite{Shanahan:2013apa}, the MIT bag model with corrections from the one-gluon-exchange and meson-cloud effects were employed to evaluate the octet baryons spin fractions carried by the valence quarks, and the obtained results were $42.6\%$, $58.9\%$ and $65.2\%$ for the $\Sigma$, $\Lambda$ and $\Xi$ hyperons, respectively. And in a very recent paper~\cite{Suh:2022atr}, the quark spin content of $SU(3)$ light and singly heavy baryons were investigated using a pion mean-field approach or the chiral quark-soliton model, flavor decomposition of the axial charges of the baryons were studied explicitly, the numerical results of the singlet axial charge for nucleon are comparable with the present results in Set II model, within $1\sigma$, while those for the other octet baryons are smaller than the present results. And it's predicted that the composite quark spin should contribute about $44\%$, $46\%$, $42\%$ and $41\%$ to spin of nucleon, $\Sigma$, $\Lambda$ and $\Xi$, respectively, these values are also close to the present results in Set II.

%%%%%%%%%%%%%%%%%%%%%%%%%%%%%%% Sec. IV %%%%%%%%%%%%%%%%%%%%%%%%%%%%%%%%

\section{Summary}
\label{conc}

To summarize, in present work, we investigate the quark orbital angular momentum of the ground octet baryons employing the extended chiral constituent quark model, in which the compact pentaquark higher Fock components in baryons are taken into account. The probabilities of the higher Fock components are determined by fitting the data for the sea flavor asymmetry of the proton.

In present model, there are two sets of wave functions for the pentaquark higher Fock components in the ground octet baryons. These two sets of wave functions could yield very different probability amplitudes of the pentaquark components, as well different quark orbital angular momentum of the octet baryons. Our numerical results show that the quark angular momentum of the nucleon should be about $0.15-0.3$, namely, the quark angular momentum could contribute $30-60\%$ to the nucleon spin. And the contributions of the intrinsic sea quark spin to the nucleon spin are not large. 

For the $\Lambda$, $\Sigma$, and $\Xi$ hyperons, it's very similar to the nucleon, the spin of the intrinsic sea quark contributes less than $10\%$ to their total spin. And the obtained quark angular momentum are in the range $0.10$-$0.28$, which are a little smaller than that of the nucleon. And the corresponding total spin arisen from the quark orbital angular momentum is about $25 \sim 50\%$.

%%%%%%%%%%%%%%%%%%%%%%%%%%%%% acknowledgments %%%%%%%%%%%%%%%%%%%%%%%%

\begin{acknowledgments}

This work is partly supported by the Chongqing Natural Science
Foundation under Project No. cstc2021jcyj-msxmX0078, and
No. cstc2019jcyj-msxmX0409,
and the National Natural Science Foundation of China under Grant Nos.
12075288, 12075133, 11735003, 11961141012 and 11835015. It is also
supported by
the Youth Innovation Promotion Association CAS, Taishan
Scholar Project of Shandong Province (Grant No.tsqn202103062),
the Higher Educational Youth Innovation Science and Technology
Program Shandong Province (Grant No. 2020KJJ004).

\end{acknowledgments}

%%%%%%%%%%%%%%%%%%%%%%%%%% References %%%%%%%%%%%%%%%%%%%%%%%%%%%%%%%%%%%


\begin{thebibliography}{99}

%%%%%%%%%%%%%%%%%%%%  Traditional quark model %%%%%%%%%%%%%%%%%%%%%%%%%%%%%%

%\cite{Capstick:2000qj}
\bibitem{Capstick:2000qj}
S.~Capstick and W.~Roberts,
%``Quark models of baryon masses and decays,''
Prog. Part. Nucl. Phys. \textbf{45} (2000), S241-S331
%doi:10.1016/S0146-6410(00)00109-5
[arXiv:nucl-th/0008028 [nucl-th]].
%436 citations counted in INSPIRE as of 23 Oct 2022

%\cite{Glozman:1995fu}
\bibitem{Glozman:1995fu}
L.~Y.~Glozman and D.~O.~Riska,
%``The Spectrum of the nucleons and the strange hyperons and chiral dynamics,''
Phys. Rept. \textbf{268}, 263-303 (1996)
%doi:10.1016/0370-1573(95)00062-3
[arXiv:hep-ph/9505422 [hep-ph]].
%692 citations counted in INSPIRE as of 15 Apr 2022

%%%%%%%%%%% EMC experiments%%%%%%%%%%%%%%%%%%%%%%%%%%%%%%%%%%%%%%%%%%%%%%%%%%%%%

\bibitem{Ashman:1987hv}
  J.~Ashman {\it et al.} [European Muon Collaboration],
  %``A Measurement of the Spin Asymmetry and Determination of the Structure Function g(1) in Deep Inelastic Muon-the proton Scattering,''
  Phys.\ Lett.\ B {\bf 206}, 364 (1988).

%\cite{Ashman:1989ig}
\bibitem{Ashman:1989ig}
  J.~Ashman {\it et al.} [European Muon Collaboration],
 % ``An Investigation of the Spin Structure of the Proton in Deep Inelastic Scattering of Polarized Muons on Polarized Protons,''
  Nucl.\ Phys.\ B {\bf 328}, 1 (1989).
%  doi:10.1016/0550-3213(89)90089-8
  %%CITATION = doi:10.1016/0550-3213(89)90089-8;%%
  %1806 citations counted in INSPIRE as of 18 Apr 2021

%%%%%%%%%%%% spin experiments %%%%%%%%%%%%%%%%%%%%%%%%%%%%%%%%%%%%%%%%

%\cite{Anthony:2000fn}
\bibitem{Anthony:2000fn}
P.~L.~Anthony {\it et al.} [E155 Collaboration],
%Measurements of the $Q^{2}$ dependence of the proton and neutron spin structure functions %$g_{1}^{p}$ and $g_{1}^n$,
  Phys.\ Lett.\ B {\bf 493}, 19 (2000).

%\cite{COMPASS:2006mhr}
\bibitem{COMPASS:2006mhr}
V.~Y.~Alexakhin \textit{et al.} [COMPASS],
%``The Deuteron Spin-dependent Structure Function g1(d) and its First Moment,''
Phys. Lett. B \textbf{647}, 8-17 (2007)
%doi:10.1016/j.physletb.2006.12.076
[arXiv:hep-ex/0609038 [hep-ex]].
%404 citations counted in INSPIRE as of 12 Apr 2022

%\cite{COMPASS:2010wkz}
\bibitem{COMPASS:2010wkz}
M.~G.~Alekseev \textit{et al.} [COMPASS],
%``The Spin-dependent Structure Function of the Proton $g_1^p$ and a Test of the Bjorken Sum Rule,''
Phys. Lett. B \textbf{690}, 466-472 (2010)
%doi:10.1016/j.physletb.2010.05.069
[arXiv:1001.4654 [hep-ex]].
%193 citations counted in INSPIRE as of 12 Apr 2022

%\cite{COMPASS:2015mhb}
\bibitem{COMPASS:2015mhb}
C.~Adolph \textit{et al.} [COMPASS],
%``The spin structure function $g_1^{\rm p}$ of the proton and a test of the %Bjorken sum rule,''
Phys. Lett. B \textbf{753}, 18-28 (2016)
%doi:10.1016/j.physletb.2015.11.064
[arXiv:1503.08935 [hep-ex]].
%107 citations counted in INSPIRE as of 12 Apr 2022

%\cite{COMPASS:2016jwv}
\bibitem{COMPASS:2016jwv}
C.~Adolph \textit{et al.} [COMPASS],
%``Final COMPASS results on the deuteron spin-dependent structure function %$g_1^{\rm d}$ and the Bjorken sum rule,''
Phys. Lett. B \textbf{769}, 34-41 (2017)
%doi:10.1016/j.physletb.2017.03.018
[arXiv:1612.00620 [hep-ex]].
%38 citations counted in INSPIRE as of 12 Apr 2022

%\cite{COMPASS:2017hef}
\bibitem{COMPASS:2017hef}
M.~Aghasyan \textit{et al.} [COMPASS],
%``Longitudinal double-spin asymmetry $A_1^{\rm p}$ and spin-dependent structure %function $g_1^{\rm p}$ of the proton at small values of $x$ and $Q^2$,''
Phys. Lett. B \textbf{781}, 464-472 (2018)
%doi:10.1016/j.physletb.2018.03.044
[arXiv:1710.01014 [hep-ex]].
%13 citations counted in INSPIRE as of 12 Apr 2022

%\cite{Deur:2021klh}
\bibitem{Deur:2021klh}
A.~Deur, J.~P.~Chen, S.~E.~Kuhn, C.~Peng, M.~Ripani, V.~Sulkosky, K.~Adhikari, M.~Battaglieri, V.~D.~Burkert and G.~D.~Cates, \textit{et al.}
%``Experimental study of the behavior of the Bjorken sum at very low Q2,''
Phys. Lett. B \textbf{825}, 136878 (2022)
%doi:10.1016/j.physletb.2022.136878
[arXiv:2107.08133 [nucl-ex]].
%3 citations counted in INSPIRE as of 12 Apr 2022

%%%%%%%%%%%%%%%%%%% Nucleon spin gauge theory %%%%%%%%%%%%%%%%%%%%%%%%%%%

%\cite{Ji:1996ek}
\bibitem{Ji:1996ek}
X.~D.~Ji,
%``Gauge-Invariant Decomposition of Nucleon Spin,''
Phys. Rev. Lett. \textbf{78} (1997), 610-613
%doi:10.1103/PhysRevLett.78.610
[arXiv:hep-ph/9603249 [hep-ph]].
%2019 citations counted in INSPIRE as of 23 Oct 2022

%\cite{Balitsky:1997rs}
\bibitem{Balitsky:1997rs}
I.~Balitsky and X.~D.~Ji,
%``How much of the nucleon spin is carried by glue?,''
Phys. Rev. Lett. \textbf{79} (1997), 1225-1228
%doi:10.1103/PhysRevLett.79.1225
[arXiv:hep-ph/9702277 [hep-ph]].
%47 citations counted in INSPIRE as of 23 Oct 2022

%\cite{Ji:1997pf}
\bibitem{Ji:1997pf}
X.~D.~Ji,
%``Lorentz symmetry and the internal structure of the nucleon,''
Phys. Rev. D \textbf{58} (1998), 056003
%doi:10.1103/PhysRevD.58.056003
[arXiv:hep-ph/9710290 [hep-ph]].
%61 citations counted in INSPIRE as of 23 Oct 2022

%\cite{Chen:2008ag}
\bibitem{Chen:2008ag}
X.~S.~Chen, X.~F.~Lu, W.~M.~Sun, F.~Wang and T.~Goldman,
%``Spin and orbital angular momentum in gauge theories: Nucleon spin structure and multipole radiation revisited,''
Phys. Rev. Lett. \textbf{100} (2008), 232002
%doi:10.1103/PhysRevLett.100.232002
[arXiv:0806.3166 [hep-ph]].
%175 citations counted in INSPIRE as of 23 Oct 2022

%%%%%%%%%%%%%%%%%%%% nucleon spin lqcd %%%%%%%%%%%%%%%%%%%%%%%%%%%%%%%%%%%%

%\cite{Aoki:1996pi}
\bibitem{Aoki:1996pi}
S.~Aoki, M.~Doui, T.~Hatsuda and Y.~Kuramashi,
%``Tensor charge of the nucleon in lattice QCD,''
Phys.\ Rev.\ D {\bf 56}, 433 (1997)
%doi:10.1103/PhysRevD.56.433
%[hep-lat/9608115].
%%CITATION = doi:10.1103/PhysRevD.56.433;%%
%97 citations counted in INSPIRE as of 17 Mar 2021

%\cite{Hagler:2007xi}
\bibitem{Hagler:2007xi}
P.~Hagler {\it et al.} [LHPC Collaboration],
%``Nucleon Generalized Parton Distributions from Full Lattice QCD,''
Phys.\ Rev.\ D {\bf 77}, 094502 (2008)
%doi:10.1103/PhysRevD.77.094502
[arXiv:0705.4295 [hep-lat]].
%%CITATION = doi:10.1103/PhysRevD.77.094502;%%
%320 citations counted in INSPIRE as of 17 Mar 2021

%\cite{QCDSF:2011aa}
\bibitem{QCDSF:2011aa}
G.~S.~Bali {\it et al.} [QCDSF Collaboration],
%``Strangeness Contribution to the Proton Spin from Lattice QCD,''
Phys.\ Rev.\ Lett.\  {\bf 108}, 222001 (2012)
%doi:10.1103/PhysRevLett.108.222001
[arXiv:1112.3354 [hep-lat]].
%%CITATION = doi:10.1103/PhysRevLett.108.222001;%%
%104 citations counted in INSPIRE as of 17 Mar 2021

%\cite{Yang:2016plb}
\bibitem{Yang:2016plb}
  Y.~B.~Yang, R.~S.~Sufian, A.~Alexandru, T.~Draper, M.~J.~Glatzmaier, K.~F.~Liu and Y.~Zhao,
 % ``Glue Spin and Helicity in the Proton from Lattice QCD,''
  Phys.\ Rev.\ Lett.\  {\bf 118}, no. 10, 102001 (2017)
  %doi:10.1103/PhysRevLett.118.102001
  [arXiv:1609.05937 [hep-ph]].
  %%CITATION = doi:10.1103/PhysRevLett.118.102001;%%
  %48 citations counted in INSPIRE as of 19 Mar 2021

%\cite{Alexandrou:2017oeh}
\bibitem{Alexandrou:2017oeh}
  C.~Alexandrou, M.~Constantinou, K.~Hadjiyiannakou, K.~Jansen, C.~Kallidonis, G.~Koutsou, A.~Vaquero Avil\'{e}s-Casco and C.~Wiese,
  %``Nucleon Spin and Momentum Decomposition Using Lattice QCD Simulations,''
  Phys.\ Rev.\ Lett.\  {\bf 119}, no. 14, 142002 (2017)
 % doi:10.1103/PhysRevLett.119.142002
  [arXiv:1706.02973 [hep-lat]].
  %%CITATION = doi:10.1103/PhysRevLett.119.142002;%%
  %80 citations counted in INSPIRE as of 23 Mar 2021

  %\cite{Yamanaka:2018uud}
\bibitem{Yamanaka:2018uud}
  N.~Yamanaka {\it et al.} [JLQCD Collaboration],
  %``Nucleon charges with dynamical overlap fermions,''
  Phys.\ Rev.\ D {\bf 98}, no. 5, 054516 (2018)
 % doi:10.1103/PhysRevD.98.054516
  [arXiv:1805.10507 [hep-lat]].
  %%CITATION = doi:10.1103/PhysRevD.98.054516;%%
  %53 citations counted in INSPIRE as of 23 Mar 2021
  
 %\cite{Yang:2019dha}
\bibitem{Yang:2019dha}
Y.~B.~Yang [Xqcd],
%``A Lattice Story of Proton Spin,''
PoS \textbf{LATTICE2018} (2019), 017
%doi:10.22323/1.334.0017
[arXiv:1904.04138 [hep-lat]].
%5 citations counted in INSPIRE as of 23 Oct 2022 

  %\cite{RQCD:2019jai}
\bibitem{RQCD:2019jai}
G.~S.~Bali \textit{et al.} [RQCD],
%``Nucleon axial structure from lattice QCD,''
JHEP \textbf{05}, 126 (2020)
%doi:10.1007/JHEP05(2020)126
[arXiv:1911.13150 [hep-lat]].
%39 citations counted in INSPIRE as of 12 Apr 2022

%%%%%%%%%%%%%%%%%% nucleon-chiPT%%%%%%%%%%%%%%%%%%%%%%%%%%%%%%%%%%%%%

%\cite{Chen:2001pva}
\bibitem{Chen:2001pva}
J.~W.~Chen and X.~d.~Ji,
%``Leading chiral contributions to the spin structure of the proton,''
Phys. Rev. Lett. \textbf{88}, 052003 (2002)
%doi:10.1103/PhysRevLett.88.052003
[arXiv:hep-ph/0111048 [hep-ph]].
%67 citations counted in INSPIRE as of 12 Apr 2022

%\cite{Dorati:2007bk}
\bibitem{Dorati:2007bk}
M.~Dorati, T.~A.~Gail and T.~R.~Hemmert,
%``Chiral perturbation theory and the first moments of the generalized parton distributions in a nucleon,''
Nucl. Phys. A \textbf{798}, 96-131 (2008)
%doi:10.1016/j.nuclphysa.2007.10.012
[arXiv:nucl-th/0703073 [nucl-th]].
%86 citations counted in INSPIRE as of 12 Apr 2022

%\cite{Lensky:2014dda}
\bibitem{Lensky:2014dda}
V.~Lensky, J.~M.~Alarc\'on and V.~Pascalutsa,
%``Moments of nucleon structure functions at next-to-leading order in baryon chiral perturbation theory,''
Phys. Rev. C \textbf{90}, no.5, 055202 (2014)
%doi:10.1103/PhysRevC.90.055202
[arXiv:1407.2574 [hep-ph]].
%55 citations counted in INSPIRE as of 12 Apr 2022

  %\cite{Li:2015exr}
\bibitem{Li:2015exr}
  H.~Li, P.~Wang, D.~B.~Leinweber and A.~W.~Thomas,
%  ``Spin of the proton in chiral effective field theory,''
  Phys.\ Rev.\ C {\bf 93}, no. 4, 045203 (2016)
%  doi:10.1103/PhysRevC.93.045203
  [arXiv:1512.02354 [hep-ph]].
  %%CITATION = doi:10.1103/PhysRevC.93.045203;%%
  %9 citations counted in INSPIRE as of 23 Mar 2021

%%%%%%%%%%%%%%%%% Nucleon-spin-others %%%%%%%%%%%%%%%%%%%%%%%%%%%%%%%%%%%%

%\cite{Brodsky:1988ip}
\bibitem{Brodsky:1988ip}
S.~J.~Brodsky, J.~R.~Ellis and M.~Karliner,
%``Chiral Symmetry and the Spin of the Proton,''
Phys. Lett. B \textbf{206} (1988), 309-315
%doi:10.1016/0370-2693(88)91511-0
%585 citations counted in INSPIRE as of 23 Oct 2022

%\cite{Qing:1997th}
\bibitem{Qing:1997th}
D.~Qing, X.~S.~Chen and F.~Wang,
%``Spin content of the nucleon in a valence and sea quark mixing model,''
Phys. Rev. C \textbf{57} (1998), R31-R34
%doi:10.1103/PhysRevC.57.R31
[arXiv:hep-ph/9712448 [hep-ph]].
%18 citations counted in INSPIRE as of 23 Oct 2022

%\cite{Qing:1998at}
\bibitem{Qing:1998at}
D.~Qing, X.~S.~Chen and F.~Wang,
%``Is nucleon spin structure inconsistent with constituent quark model?,''
Phys. Rev. D \textbf{58} (1998), 114032
%doi:10.1103/PhysRevD.58.114032
[arXiv:hep-ph/9802425 [hep-ph]].
%33 citations counted in INSPIRE as of 23 Oct 2022

%\cite{Brodsky:2000ii}
\bibitem{Brodsky:2000ii}
S.~J.~Brodsky, D.~S.~Hwang, B.~Q.~Ma and I.~Schmidt,
%``Light cone representation of the spin and orbital angular momentum of relativistic composite systems,''
Nucl. Phys. B \textbf{593} (2001), 311-335
%doi:10.1016/S0550-3213(00)00626-X
[arXiv:hep-th/0003082 [hep-th]].
%341 citations counted in INSPIRE as of 23 Oct 2022

%\cite{An:2005cj}
\bibitem{An:2005cj}
  C.~S.~An, D.~O.~Riska and B.~S.~Zou,
 % ``Strangeness spin, magnetic moment and strangeness configurations of the proton,''
  Phys.\ Rev.\ C {\bf 73}, 035207 (2006)
  %doi:10.1103/PhysRevC.73.035207
  [hep-ph/0511223].
  %%CITATION = doi:10.1103/PhysRevC.73.035207;%%
  %54 citations counted in INSPIRE as of 19 Mar 2021
  
  %\cite{Brodsky:2006ha}
\bibitem{Brodsky:2006ha}
S.~J.~Brodsky and S.~Gardner,
%``Evidence for the Absence of Gluon Orbital Angular Momentum in the Nucleon,''
Phys. Lett. B \textbf{643} (2006), 22-28
%doi:10.1016/j.physletb.2006.10.024
[arXiv:hep-ph/0608219 [hep-ph]].
%74 citations counted in INSPIRE as of 23 Oct 2022

%\cite{Barquilla-Cano:2006hku}
\bibitem{Barquilla-Cano:2006hku}
D.~Barquilla-Cano, A.~J.~Buchmann and E.~Hernandez,
%``Axial exchange currents and nucleon spin,''
Eur. Phys. J. A \textbf{27} (2006), 365-372
%doi:10.1140/epja/i2005-10270-4
[arXiv:hep-ph/0611248 [hep-ph]].
%13 citations counted in INSPIRE as of 06 Jan 2023

  %\cite{Adamuscin:2007fk}
\bibitem{Adamuscin:2007fk}
  C.~Adamuscin, E.~Tomasi-Gustafsson, E.~Santopinto and R.~Bijker,
  %``Two-component model for the axial form factor of the nucleon,''
  Phys.\ Rev.\ C {\bf 78}, 035201 (2008)
  %doi:10.1103/PhysRevC.78.035201
  [arXiv:0706.3509 [nucl-th]].
  %%CITATION = doi:10.1103/PhysRevC.78.035201;%%
  %18 citations counted in INSPIRE as of 19 Mar 2021

%\cite{Myhrer:2007cf}
\bibitem{Myhrer:2007cf}
F.~Myhrer and A.~W.~Thomas,
%``A possible resolution of the proton spin problem,''
Phys.\ Lett.\ B {\bf 663}, 302 (2008)
%doi:10.1016/j.physletb.2008.04.034
[arXiv:0709.4067 [hep-ph]].
%%CITATION = doi:10.1016/j.physletb.2008.04.034;%%
%76 citations counted in INSPIRE as of 17 Mar 2021

%\cite{Thomas:2008ga}
\bibitem{Thomas:2008ga}
A.~W.~Thomas,
%``Interplay of Spin and Orbital Angular Momentum in the Proton,''
Phys. Rev. Lett. \textbf{101} (2008), 102003
%doi:10.1103/PhysRevLett.101.102003
[arXiv:0803.2775 [hep-ph]].
%109 citations counted in INSPIRE as of 23 Oct 2022
  
  %\cite{Bijker:2009up}
\bibitem{Bijker:2009up}
R.~Bijker and E.~Santopinto,
%``Unquenched quark model for baryons: Magnetic moments, spins and orbital angular momenta,''
Phys. Rev. C \textbf{80} (2009), 065210
%doi:10.1103/PhysRevC.80.065210
[arXiv:0912.4494 [nucl-th]].
%91 citations counted in INSPIRE as of 23 Oct 2022

 %\cite{Lorce:2011kd}
\bibitem{Lorce:2011kd}
C.~Lorce and B.~Pasquini,
%``Quark Wigner Distributions and Orbital Angular Momentum,''
Phys. Rev. D \textbf{84} (2011), 014015
%doi:10.1103/PhysRevD.84.014015
[arXiv:1106.0139 [hep-ph]].
%246 citations counted in INSPIRE as of 23 Oct 2022

%%%%%%%%%%%%%%%%%%%%%%%  Reviews nucleon spin %%%%%%%%%%%%%%%%%%%%%%%%%%%%%%%%
  
%\cite{Kuhn:2008sy}
\bibitem{Kuhn:2008sy}
S.~E.~Kuhn, J.~P.~Chen and E.~Leader,
%``Spin Structure of the Nucleon - Status and Recent Results,''
Prog. Part. Nucl. Phys. \textbf{63} (2009), 1-50
%doi:10.1016/j.ppnp.2009.02.001
[arXiv:0812.3535 [hep-ph]].
%173 citations counted in INSPIRE as of 25 Oct 2022

%\cite{Burkardt:2008jw}
\bibitem{Burkardt:2008jw}
M.~Burkardt, C.~A.~Miller and W.~D.~Nowak,
%``Spin-polarized high-energy scattering of charged leptons on nucleons,''
Rept. Prog. Phys. \textbf{73} (2010), 016201
%doi:10.1088/0034-4885/73/1/016201
[arXiv:0812.2208 [hep-ph]].
%118 citations counted in INSPIRE as of 25 Oct 2022

%\cite{Leader:2013jra}
\bibitem{Leader:2013jra}
E.~Leader and C.~Lorc\'e,
%``The angular momentum controversy: What\textquoteright{}s it all about and does it matter?,''
Phys. Rept. \textbf{541} (2014) no.3, 163-248
%doi:10.1016/j.physrep.2014.02.010
[arXiv:1309.4235 [hep-ph]].
%267 citations counted in INSPIRE as of 25 Oct 2022

%\cite{Wakamatsu:2014zza}
\bibitem{Wakamatsu:2014zza}
M.~Wakamatsu,
%``Is gauge-invariant complete decomposition of the nucleon spin possible?,''
Int. J. Mod. Phys. A \textbf{29} (2014), 1430012
%doi:10.1142/S0217751X14300129
[arXiv:1402.4193 [hep-ph]].
%87 citations counted in INSPIRE as of 25 Oct 2022

%\cite{Liu:2015xha}
\bibitem{Liu:2015xha}
K.~F.~Liu and C.~Lorc\'e,
%``The Parton Orbital Angular Momentum: Status and Prospects,''
Eur. Phys. J. A \textbf{52} (2016) no.6, 160
%doi:10.1140/epja/i2016-16160-8
[arXiv:1508.00911 [hep-ph]].
%37 citations counted in INSPIRE as of 25 Oct 2022

%\cite{Deur:2018roz}
\bibitem{Deur:2018roz}
  A.~Deur, S.~J.~Brodsky and G.~F.~De T\'{e}ramond,
 % ``The Spin Structure of the Nucleon,''
  Rept.\ Prog.\ Phys.\  {\bf 82}, no. 076201 (2019)
 % doi:10.1088/1361-6633/ab0b8f
  [arXiv:1807.05250 [hep-ph]].
  %%CITATION = doi:10.1088/1361-6633/ab0b8f;%%
  %34 citations counted in INSPIRE as of 19 Mar 2021

%\cite{Ji:2020ena}
\bibitem{Ji:2020ena}
X.~Ji, F.~Yuan and Y.~Zhao,
%``What we know and what we don\textquoteright{}t know about the proton spin after 30 years,''
Nature Rev. Phys. \textbf{3}, no.1, 27-38 (2021)
%doi:10.1038/s42254-020-00248-4
[arXiv:2009.01291 [hep-ph]].
%27 citations counted in INSPIRE as of 12 Apr 2022
 
%%%%%%%%%%%%%%%%%% sea flavor asymmetry %%%%%%%%%%%%%%%%%%%%%%%%%%%%%%%%%%%%%%%%

%\cite{NuSea:2001idv}
\bibitem{NuSea:2001idv}
R.~S.~Towell \textit{et al.} [NuSea],
%``Improved measurement of the anti-d / anti-u asymmetry in the nucleon sea,''
Phys. Rev. D \textbf{64}, 052002 (2001)
%doi:10.1103/PhysRevD.64.052002
[arXiv:hep-ex/0103030 [hep-ex]].
%501 citations counted in INSPIRE as of 12 Apr 2022

%\cite{SeaQuest:2021zxb}
\bibitem{SeaQuest:2021zxb}
J.~Dove \textit{et al.} [SeaQuest],
%``The asymmetry of antimatter in the proton,''
Nature \textbf{590}, no.7847, 561-565 (2021)
%doi:10.1038/s41586-021-03282-z
[arXiv:2103.04024 [hep-ph]].
%32 citations counted in INSPIRE as of 12 Apr 2022

%%%%%%%%%%%%%%%%%%%%%%%  meson cloud model %%%%%%%%%%%%%%%%%%%%%%%%%%%%%%%%%%

%\cite{Ericson:1983um}
\bibitem{Ericson:1983um}
M.~Ericson and A.~W.~Thomas,
%``Pionic Corrections and the EMC Enhancement of the Sea in Iron,''
Phys. Lett. B \textbf{128} (1983), 112-116
%doi:10.1016/0370-2693(83)90085-0
%361 citations counted in INSPIRE as of 23 Oct 2022

%\cite{Thomas:1983fh}
\bibitem{Thomas:1983fh}
A.~W.~Thomas,
%``A Limit on the Pionic Component of the Nucleon Through SU(3) Flavor Breaking in the Sea,''
Phys. Lett. B \textbf{126} (1983), 97-100
%doi:10.1016/0370-2693(83)90026-6
%338 citations counted in INSPIRE as of 23 Oct 2022

%\cite{Henley:1990kw}
\bibitem{Henley:1990kw}
E.~M.~Henley and G.~A.~Miller,
%``Excess of anti-D over anti-U in the proton sea quark distribution,''
Phys. Lett. B \textbf{251} (1990), 453-454
%doi:10.1016/0370-2693(90)90735-O
%214 citations counted in INSPIRE as of 23 Oct 2022

%\cite{Brodsky:1996hc}
\bibitem{Brodsky:1996hc}
S.~J.~Brodsky and B.~Q.~Ma,
%``The Quark / anti-quark asymmetry of the nucleon sea,''
Phys. Lett. B \textbf{381} (1996), 317-324
%doi:10.1016/0370-2693(96)00597-7
[arXiv:hep-ph/9604393 [hep-ph]].
%278 citations counted in INSPIRE as of 23 Oct 2022

%\cite{Shao:2010wq}
\bibitem{Shao:2010wq}
L.~Shao, Y.~J.~Zhang and B.~Q.~Ma,
%``Sea quark contents of octet baryons,''
Phys. Lett. B \textbf{686} (2010), 136-140
%doi:10.1016/j.physletb.2010.02.049
[arXiv:1002.4747 [hep-ph]].
%24 citations counted in INSPIRE as of 23 Oct 2022

%%%%%%%%%%%%%%%%%%% ECQM-applications %%%%%%%%%%%%%%%%%%%%%%%%

%\cite{Zou:2005xy}
\bibitem{Zou:2005xy}
B.~S.~Zou and D.~O.~Riska,
%``The s anti-s component of the proton and the strangeness magnetic moment,''
Phys. Rev. Lett. \textbf{95} (2005), 072001
%doi:10.1103/PhysRevLett.95.072001
[arXiv:hep-ph/0502225 [hep-ph]].
%95 citations counted in INSPIRE as of 23 Oct 2022

%\cite{Riska:2005bh}
\bibitem{Riska:2005bh}
D.~O.~Riska and B.~S.~Zou,
%``The Strangeness form-factors of the proton,''
Phys. Lett. B \textbf{636} (2006), 265-269
%doi:10.1016/j.physletb.2006.03.065
[arXiv:nucl-th/0512102 [nucl-th]].
%33 citations counted in INSPIRE as of 23 Oct 2022

%\cite{An:2012kj}
\bibitem{An:2012kj}
C.~S.~An and B.~Saghai,
%``Sea flavor content of octet baryons and intrinsic five-quark Fock states,''
Phys.\ Rev.\ C {\bf 85}, 055203 (2012)
%doi:10.1103/PhysRevC.85.055203
[arXiv:1204.0300 [hep-ph]].
%%CITATION = doi:10.1103/PhysRevC.85.055203;%%
%14 citations counted in INSPIRE as of 17 Mar 2021

%\cite{An:2014aea}
\bibitem{An:2014aea}
C.~S.~An and B.~Saghai,
%``Pion- and strangeness-baryon $\sigma$ terms in the extended chiral constituent quark model,''
Phys. Rev. D \textbf{92}, no.1, 014002 (2015)
%doi:10.1103/PhysRevD.92.014002
[arXiv:1404.2389 [hep-ph]].
%7 citations counted in INSPIRE as of 12 Apr 2022

%\cite{Duan:2016rkr}
\bibitem{Duan:2016rkr}
S.~Duan, C.~S.~An and B.~Saghai,
%``Intrinsic charm content of the nucleon and charmness-nucleon sigma term,''
Phys. Rev. D \textbf{93}, no.11, 114006 (2016)
%doi:10.1103/PhysRevD.93.114006
[arXiv:1606.02000 [hep-ph]].
%12 citations counted in INSPIRE as of 12 Apr 2022

%\cite{Li:2005jn}
\bibitem{Li:2005jn}
Q.~B.~Li and D.~O.~Riska,
%``Five-quark components in Delta(1232) ---\ensuremath{>} N pi decay,''
Phys. Rev. C \textbf{73} (2006), 035201
%doi:10.1103/PhysRevC.73.035201
[arXiv:nucl-th/0507008 [nucl-th]].
%35 citations counted in INSPIRE as of 23 Oct 2022

%\cite{Li:2005jb}
\bibitem{Li:2005jb}
Q.~B.~Li and D.~O.~Riska,
%``The Role of five-quark components in gamma decay of the Delta(1232),''
Nucl. Phys. A \textbf{766} (2006), 172-182
%doi:10.1016/j.nuclphysa.2005.11.020
[arXiv:nucl-th/0511053 [nucl-th]].
%24 citations counted in INSPIRE as of 23 Oct 2022

%\cite{Li:2006nm}
\bibitem{Li:2006nm}
Q.~B.~Li and D.~O.~Riska,
%``The Role of q anti-q components in the N(1440) resonance,''
Phys. Rev. C \textbf{74} (2006), 015202
%doi:10.1103/PhysRevC.74.015202
[arXiv:nucl-th/0605076 [nucl-th]].
%42 citations counted in INSPIRE as of 23 Oct 2022

%\cite{An:2008xk}
\bibitem{An:2008xk}
C.~S.~An and B.~S.~Zou,
%``The Role of the qqqq anti-q components in the electromagnetic transition gamma*N ---\ensuremath{>} N(1535),''
Eur. Phys. J. A \textbf{39} (2009), 195-204
%doi:10.1140/epja/i2008-10698-x
[arXiv:0802.3996 [nucl-th]].
%55 citations counted in INSPIRE as of 23 Oct 2022

%\cite{An:2009uv}
\bibitem{An:2009uv}
C.~S.~An and B.~S.~Zou,
%``Strong decays of N*(1535) in an extended chiral quark model,''
Sci. China G \textbf{52} (2009), 1452-1457
%doi:10.1007/s11433-009-0199-6
[arXiv:0910.4452 [nucl-th]].
%14 citations counted in INSPIRE as of 23 Oct 2022

%\cite{An:2011sb}
\bibitem{An:2011sb}
C.~An and B.~Saghai,
%``Strong decay of low-lying $S_{11}$ and $D_{13}$ nucleon resonances to pseudoscalar mesons and octet baryons,''
Phys. Rev. C \textbf{84} (2011), 045204
%doi:10.1103/PhysRevC.84.045204
[arXiv:1108.3282 [nucl-th]].
%23 citations counted in INSPIRE as of 23 Oct 2022

%\cite{An:2019tld}
\bibitem{An:2019tld}
C.~S.~An and B.~Saghai,
%``Orbital angular momentum of the proton and intrinsic five-quark Fock states,''
Phys. Rev. D \textbf{99}, no.9, 094039 (2019)
%doi:10.1103/PhysRevD.99.094039
[arXiv:1905.05330 [hep-ph]].
%1 citations counted in INSPIRE as of 12 Apr 2022

%\cite{Wang:2021ild}
\bibitem{Wang:2021ild}
J.~B.~Wang, G.~Li, C.~S.~An and J.~J.~Xie,
%``Axial charges of the proton within an extended chiral constituent quark model,''
Phys. Rev. D \textbf{103}, no.11, 114018 (2021)
%doi:10.1103/PhysRevD.103.114018
[arXiv:2106.00866 [hep-ph]].
%1 citations counted in INSPIRE as of 12 Apr 2022

%\cite{Qi:2022sus}
\bibitem{Qi:2022sus}
R.~Qi, J.~B.~Wang, G.~Li, C.~S.~An, C.~R.~Deng and J.~J.~Xie,
%``Investigations on the flavor-dependent axial charges of the octet baryons,''
Phys. Rev. C \textbf{105} (2022) no.6, 065204
%doi:10.1103/PhysRevC.105.065204
[arXiv:2205.14419 [hep-ph]].
%1 citations counted in INSPIRE as of 23 Oct 2022

%%%%%%%%%%%%%%%%%%%%%%%%%  PDG %%%%%%%%%%%%%%%%%%%%%%%%%%%%%%%%%%%%%%%%%%%%%%%%%

%\cite{ParticleDataGroup:2020ssz}
\bibitem{ParticleDataGroup:2020ssz}
P.~A.~Zyla \textit{et al.} [Particle Data Group],
``Review of Particle Physics,''
PTEP \textbf{2020}, no.8, 083C01 (2020)
%doi:10.1093/ptep/ptaa104
%3327 citations counted in INSPIRE as of 16 Apr 202


%%%%%%%%%%%%%%%%%%%%%%%%%%%%  3P0 model %%%%%%%%%%%%%%%%%%%%%%%%%%%%%%%%%%%%%%

%\cite{LeYaouanc:1972vsx}
\bibitem{LeYaouanc:1972vsx}
A.~Le Yaouanc, L.~Oliver, O.~Pene and J.~C.~Raynal,
%``Naive quark pair creation model of strong interaction vertices,''
Phys. Rev. D \textbf{8} (1973), 2223-2234
%doi:10.1103/PhysRevD.8.2223
%691 citations counted in INSPIRE as of 25 Oct 2022

%\cite{LeYaouanc:1973ldf}
\bibitem{LeYaouanc:1973ldf}
A.~Le Yaouanc, L.~Oliver, O.~Pene and J.~C.~Raynal,
%``Naive quark pair creation model and baryon decays,''
Phys. Rev. D \textbf{9} (1974), 1415-1419
%doi:10.1103/PhysRevD.9.1415
%341 citations counted in INSPIRE as of 25 Oct 2022

%%%%%%%%%%%%%%%%%%  BHPS model %%%%%%%%%%%%%%%%%%%%%%%%%%%%%%%%%%%%%%%%%%%%%%%%
%\cite{Chang:2011vx}
\bibitem{Chang:2011vx}
  W.~C.~Chang and J.~C.~Peng,
  ``Flavor Asymmetry of the Nucleon Sea and the Five-Quark Components of the Nucleons,''
  Phys.\ Rev.\ Lett.\  {\bf 106}, 252002 (2011)
  %doi:10.1103/PhysRevLett.106.252002
  [arXiv:1102.5631 [hep-ph]].
  %%CITATION = doi:10.1103/PhysRevLett.106.252002;%%
  %57 citations counted in INSPIRE as of 25 Mar 2021
  
  %\cite{Chang:2011du}
\bibitem{Chang:2011du}
W.~C.~Chang and J.~C.~Peng,
%``Extraction of Various Five-Quark Components of the Nucleons,''
Phys. Lett. B \textbf{704} (2011), 197-200
%doi:10.1016/j.physletb.2011.08.077
[arXiv:1105.2381 [hep-ph]].
%49 citations counted in INSPIRE as of 23 Oct 2022

%\cite{Chang:2014jba}
\bibitem{Chang:2014jba}
W.~C.~Chang and J.~C.~Peng,
%``Flavor Structure of the Nucleon Sea,''
Prog. Part. Nucl. Phys. \textbf{79} (2014), 95-135
%doi:10.1016/j.ppnp.2014.08.002
[arXiv:1406.1260 [hep-ph]].
%71 citations counted in INSPIRE as of 23 Oct 2022

%\cite{Brodsky:1980pb}
\bibitem{Brodsky:1980pb}
S.~J.~Brodsky, P.~Hoyer, C.~Peterson and N.~Sakai,
%``The Intrinsic Charm of the Proton,''
Phys. Lett. B \textbf{93} (1980), 451-455
%doi:10.1016/0370-2693(80)90364-0
%826 citations counted in INSPIRE as of 24 Oct 2022

%%%%%%%%%%%%%%%%%%%%%%%%%%%%%%%%%  PCAC  %%%%%%%%%%%%%%%%%%%%%%%%%%%%%%%%%

%\cite{Barquilla-Cano:2002btm}
\bibitem{Barquilla-Cano:2002btm}
D.~Barquilla-Cano, A.~J.~Buchmann and E.~Hernandez,
%``Partial conservation of axial current and axial exchange currents in the nucleon,''
Nucl. Phys. A \textbf{714} (2003), 611-631
%doi:10.1016/S0375-9474(02)01389-1
[arXiv:nucl-th/0204067 [nucl-th]].
%13 citations counted in INSPIRE as of 06 Jan 2023

%\cite{Barquilla-Cano:2003qmv}
\bibitem{Barquilla-Cano:2003qmv}
D.~Barquilla-Cano, E.~Hernandez and A.~J.~Buchmann,
%``Axial exchange currents and the spin content of the nucleon in a nonrelativistic chiral quark model,''
Nucl. Phys. A \textbf{721} (2003), 429-432
%doi:10.1016/S0375-9474(03)01089-3
%3 citations counted in INSPIRE as of 06 Jan 2023

%%%%%%%%%%%%%%%%%%%%%%%%%%%%%%%%%  octet spin %%%%%%%%%%%%%%%%%%%%%%%%%%%%%%%%%

%\cite{Buchmann:2010pd}
\bibitem{Buchmann:2010pd}
A.~J.~Buchmann and E.~M.~Henley,
%``Spin of ground state baryons,''
Phys. Rev. D \textbf{83} (2011), 096011
%doi:10.1103/PhysRevD.83.096011
[arXiv:1011.0139 [hep-ph]].
%9 citations counted in INSPIRE as of 06 Jan 2023

%\cite{Buchmann:2014ima}
\bibitem{Buchmann:2014ima}
A.~J.~Buchmann and E.~M.~Henley,
%``Three-Quark Currents and Baryon Spin,''
Few Body Syst. \textbf{55} (2014), 749-752
%doi:10.1007/s00601-014-0864-9
[arXiv:1408.1311 [hep-ph]].
%2 citations counted in INSPIRE as of 06 Jan 2023

%\cite{Shanahan:2013apa}
\bibitem{Shanahan:2013apa}
P.~E.~Shanahan, A.~W.~Thomas, K.~Tsushima, R.~D.~Young and F.~Myhrer,
%``Octet Spin Fractions and the Proton Spin Problem,''
Phys. Rev. Lett. \textbf{110} (2013) no.20, 202001
%doi:10.1103/PhysRevLett.110.202001
[arXiv:1302.6300 [nucl-th]].
%18 citations counted in INSPIRE as of 24 Oct 2022

%\cite{Suh:2022atr}
\bibitem{Suh:2022atr}
J.~M.~Suh, J.~Y.~Kim, G.~S.~Yang and H.~C.~Kim,
%``Quark spin content of SU(3) light and singly heavy baryons,''
Phys. Rev. D \textbf{106} (2022) no.5, 054032
%doi:10.1103/PhysRevD.106.054032
[arXiv:2208.04447 [hep-ph]].
%2 citations counted in INSPIRE as of 25 Oct 2022

\end{thebibliography}
\end{document}